\documentstyle[12pt]{article}
\catcode`\@=11

\def\gut{{\rm GUT}}
\def\str{{\rm str}}
\def\sm{SU_c(3)\times SU_L(2)\times U_Y(1)}
\def\phi5{\Phi_5}
\def\phi5b{\overline{\Phi_5}}
\def\sige{\Sigma_8}

\def\sigt{\Sigma_3}

\def\nphi{n^{}_\Phi}
\def\nsig{n^{}_\Sigma}
\input{epsf}
\ifx\epsffile\undefined
\newlength{\epsfysize}
\def\epsffile#1#2#3#4]#5{}
\fi

\begin{document}
\begin{titlepage}
\samepage{
\setcounter{page}{1}

\begin{flushright}
{\small 
MADPH-98-1053\\
hep-ph/9804228\\
April, 1998\\}
\end{flushright}

\begin{center}

{\Large{\bf Adjoint Messengers and Perturbative Unification\\
at the String Scale}}

\vfil
{\large{T. Han$^1$, T. Yanagida$^{1,2}$ and R.-J. Zhang$^1$\\}}
\vspace{.25in}
 {\it  $^1$Department of Physics, University of 
Wisconsin, Madison, WI 53706, USA\\
       $^2$Department of Physics, University of Tokyo, Tokyo 113, Japan}
\end{center}
\vfil

\begin{abstract}
{\rm 
We consider states in the adjoint representation of the 
Standard Model gauge group as messengers for mediation of 
supersymmetry (SUSY) breaking. These new messengers can shift the
gauge coupling unification to the string scale at 
${\cal O}(5\times 10^{17}$ GeV) if their masses are at 
${\cal O}(10^{14}$ GeV). The predicted SUSY mass spectrum 
at the electroweak scale is significantly different 
from those in other gauge-mediated or supergravity
models, resulting in robust mass relations. 
The gravitino mass is predicted to be about $1-10$ GeV. 
The heavy messenger sector could provide
a superheavy dark matter candidate.}
\end{abstract}
\vfil }
\end{titlepage}

\newcommand{\gsim}{\lower.7ex\hbox{$\;\stackrel{\textstyle>}{\sim}\;$}}
\newcommand{\lsim}{\lower.7ex\hbox{$\;\stackrel{\textstyle<}{\sim}\;$}}
\newcommand{\GeV}{{\rm GeV}}
\newcommand{\TeV}{{\rm TeV}}
\newcommand{\La}{{\Lambda}}

{\noindent\bf{Introduction:}}~
Weak scale supersymmetry (SUSY) is arguably the strongest candidate
for physics beyond the Standard Model (SM). 
One of the most attractive features is that 
the supersymmetric $SU(5)$ theory provides a 
non-trivial coupling unification for the SM gauge group
$SU_c(3)\times SU_L(2)\times U_Y(1)$,
consistent with the experimental determination for
the coupling constants at 
the electroweak scale ($M_Z$) \cite{susygut}. 
In the minimal supersymmetric extension of the Standard Model 
(MSSM) with a SUSY mass scale near 1 TeV, 
the gauge coupling unification occurs 
at a scale $M_{\gut} \approx 2\times 10^{16}$ GeV.
On the other hand, heterotic string theories generically 
predict a perturbative string unification at a scale
$M_{\str} \approx 5\times 10^{17}$ GeV. These two scales
are mysteriously close (in relative value), 
yet significantly different (in absolute value).
It is therefore extremely tempting to contemplate on
physics scenarios to reconcile these two scales \cite{strgut}. 

One of the possibilities to fulfill this idea is to 
introduce some states beyond the MSSM below the 
unification scale. The additional states modify the behavior
of the gauge coupling evolution and may lift the
unification scale from $M_\gut$ to $M_\str$.
An explicit example has been constructed by considering
adjoint representations for an
$SU_c(3)$ octet ($\Sigma_8$) plus an $SU_L(2)$ triplet 
($\Sigma_3$) \cite{bfy}. This scenario is
particularly interesting since the new states 
could naturally arise from the non-Goldstone
remnants of the Higgs multiplets $\Sigma_{24}$.

In spite of our ignorance about
the SUSY breaking mechanism at high energy scale, one
would like to at least explore how the SUSY breaking
effects have been transmitted to the observable sector
at the electroweak scale. A model with gauge mediation 
of SUSY breaking (GMSB) \cite{DN} is a simple and 
predictive version of the MSSM. In addition to the
observable sector and a SUSY breaking hidden 
sector, the model also possesses messenger fields which 
mediate the SUSY breaking to the observable fields via 
the SM gauge interactions. The {\it minimal} model has 
a pair of messengers $\Phi_5+\overline{\Phi}_5$
transforming under the $SU(5)$
representations ${\bf 5+{\overline 5}}$. 
By assumption, this minimal model of gauge-mediated
SUSY breaking (mGMSB) contains messengers as complete 
$SU(5)$ representations. This construction automatically
preserves the gauge coupling unification at $M_\gut$.

In this Letter, we propose a ``marriage'' of these two ideas:
we introduce some states beyond the MSSM below $M_\gut$ 
(incomplete representations of the GUT group)
as new messengers to mediate the SUSY breaking effects, 
and the scale of gauge coupling unification is lifted to 
$M_\str$. This scenario may
have profound theoretical implication: the gauge coupling
unification at the string scale may be intimately
connected with the gauge mediation of the SUSY breaking.
It is important to note that the introduction
of the new messengers predicts a different mass spectrum 
for SUSY particles (sparticles) from those
in the GMSB models \cite{gmsbmass} and 
in the supergravity models (SUGRA) \cite{sugramass}. 
Thus, we should be able to test this idea once 
the SUSY mass parameters
are measured at future collider experiments. 
There are also interesting cosmological consequences
in this scenario that we will discuss in the later sections.

\begin{figure}[tb]
\epsfysize=2.5in
\epsffile[-20 280 380 545]{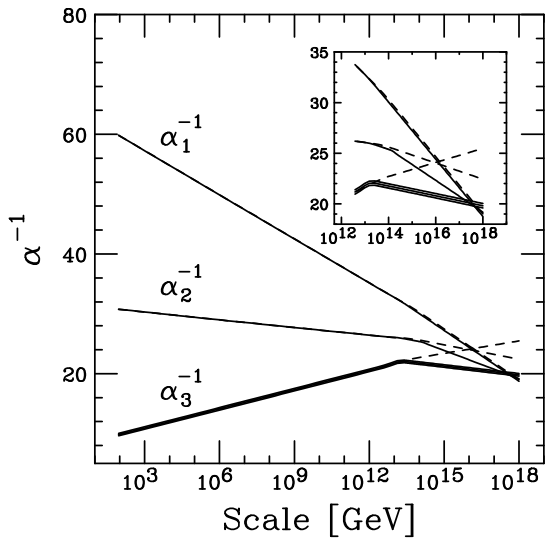}
\begin{center}
\parbox{6.0in}{
\caption[]{Evolution of the SM gauge couplings
from the electroweak scale $M_Z^{}$ to the string unification
scale $M_\str$ with a generic messenger scale
$M\approx 10^{14}$ GeV.
The solid curves show the gauge coupling
unification at $M_\str$ with the help of the adjoint
messengers. The dashes give the unification at
$M_\gut$ without the adjoint messengers, but with
$\Phi_5+{\overline \Phi}_5$ at $M$. 
The inner panel shows the blowup in the unification
region. We have taken 
$\tan\beta=2$, $\mu<0$ and $\alpha_s(M_Z)=0.118\pm 0.003$.
\label{unify}}}
\end{center}
\end{figure}

\vskip 0.1in
{\noindent \bf{Adjoint Messengers and Gauge Coupling Unification:}}~
Following the proposal in Ref.~\cite{bfy} to resolve the
string unification problem, we first introduce a pair of
new messenger fields $\Sigma_8$ and $\Sigma_3$ with
the following $\sm$ quantum number assignment
\begin{equation}
\Sigma_8: \ \ ({\bf 8,1},Y=0);\qquad
\Sigma_3: \ \ ({\bf 1,3},Y=0).
\label{numb}
\end{equation}
They are in adjoint representations and thus anomaly-free,
which will be referred as ``adjoint messengers''. 
We consider a general model which includes   
$\nphi$ pairs of $\Phi_5+\overline{\Phi}_5$ and
$\nsig$ pairs of $\sige+\sigt$ states.
The renormalization group equations (RGEs) at one-loop level
for the SM gauge couplings, 
$\alpha_i=g^2_i/4\pi \ (i=1,2,3)$, 
up to $M_\str$ are given by
\begin{eqnarray}
{d\alpha_i\over dt} = (b^{\rm MSSM}_i+N_i ){\alpha_i^2\over 2\pi},
\end{eqnarray}
where $t=\ln Q$ and 
the $b$-coefficients in the MSSM are
$b^{\rm MSSM}_{1,2,3}=+33/5,+1,-3$ respectively, and
$N_i$ the new state counting
\begin{equation}
N_1= \nphi,\ N_2=\nphi+2\nsig,\ N_3=\nphi+3\nsig
\quad({\rm above\ \Phi_5, \sige\ and\ \sigt\ threshold}\ M).
\end{equation}
Since $\Phi_5+\overline{\Phi}_5$ form complete $SU(5)$
representations, they preserve the unification at the
$\gut$ scale and their masses can be arbitrary 
between the MSSM threshold and $M_\gut$. In contrast,
the adjoint messengers can change the running behavior 
of the couplings and shift the unification
scale around depending upon the number of states
and their masses.
We find that as long as we take the same
number of states for $\Sigma_8$ and $\Sigma_3$, 
a unification can be achieved. 
This justifies our choice for a single $\nsig$. 
The evolution of the couplings from $M_Z^{}$
to $M_\str$ is illustrated in Fig.~\ref{unify},
where we have evolved the couplings at two-loop 
level, including SUSY threshold corrections at
the electroweak scale. 
The solid curves demonstrate the string
scale unification with $\nphi=\nsig=1$, and the dashes 
show the unification with $\nphi=1$ only.
To reach a successful unification at
$M_\str\approx 5\times 10^{17}$ GeV and accommodate the strong
coupling constant $\alpha_s(M_Z) = 0.118$, the masses of 
the adjoint messengers need to be \cite{twoloop}
\begin{equation}
M_8\approx 2.5\times 10^{13}\ {\rm GeV} ,
\qquad 
M_3\approx 1.2\times 10^{14}\ {\rm GeV} ,
\label{mm}
\end{equation}
for which the gauge couplings unify to
$\alpha_\gut^{}(M_\str) \approx 1/20$.
Following Eq.~(\ref{mm}), we will generically identify 
the messenger scale ($M$-scale)
for $\Phi_5+\overline{\Phi}_5$ and $\sige+\sigt$ as 
\begin{equation}
M\approx 10^{14}\ {\rm GeV}. 
\label{mscale}
\end{equation}

Perturbativity requirement for the gauge couplings up to  $M_\str$ 
leads to a bound on the numbers of the messenger
states, $\nphi, \nsig \le {\cal O}(10)$, 
for their masses $M \leq {\cal O}(10^{14}$ GeV).
Such a loose bound is due to the smaller running effects
between the rather close scales $M$ and $M_\str$. On the
other hand, many pairs of the adjoint messengers would 
push their mass scale too close to $M_\str$ for the
unification. Furthermore, to avoid a too heavy 
sparticle spectrum, especially for the gravitino 
mass ($m_{3/2}$) as we will discuss later, $\nsig=1$
is strongly favored. For concreteness, we also choose $\nphi=1$
in the rest of our studies, which can be easily generalized to
other values of $\nphi$.

\vskip 0.1in
{\noindent \bf{Predicted SUSY Mass Spectrum and Physical Consequences:}}~
In GMSB models, the messengers couple to gauge singlet 
fields $S_i$ through a superpotential
\begin{equation}
W = 
\lambda_5^{} S_5 \Phi_5 \overline{\Phi}_5 +
\lambda_8^{} S_8 \sige \sige + \lambda_3^{} S_3 \sigt \sigt .
\label{new}
\end{equation}
For simplicity, we take the singlets to be the same $S_5=S_8=S_3$, 
which acquires non-zero vacuum expectation values for both
its scalar component ($S$) and auxiliary component ($F_S$). 

One of the important features for this model 
is that all sparticle masses are determined 
by two dimensionful parameters: the messenger scale 
$M=\lambda S$ and the effective SUSY breaking scale 
$\Lambda= F_S/ S$. The gaugino and scalar soft 
masses are given, at one- and two-loop level respectively, 
by \cite{DN}
\begin{eqnarray}
\label{gm0}
  M_i(M) &\approx&  N_i {\alpha_i(M)\over 4\pi}\La \ ,
      \qquad i=1,2,3\\
  {\tilde m}^2(M) &\approx& 2 \sum_{i=1}^3 N_i  C_i 
  \left({{\alpha_i}(M)\over 4\pi}\right)^2 \La^2 \ ,
\label{gm1}
\end{eqnarray}
where $C_i$'s are $4/3$, $3/4$ 
for the fundamental representations of $SU_c(3)$, $SU_L(2)$ and $3Y^2/5$
for $U_Y(1)$. 

Equation~(\ref{gm0}) implies a gaugino mass relation
\begin{equation}
{M_1(M)\over \alpha_1(M)}:{M_2(M)\over \alpha_2(M)}:
{M_3(M)\over \alpha_3(M)}=
N_1 : N_2 : N_3\ .
\label{ratiochi}
\end{equation}
Alternatively, we can rewrite the mass ratio relation,
independent of $\nphi$ and $\nsig$, as
\begin{equation}
\left( {M_2(M)\over \alpha_2(M)}- {M_1(M)\over \alpha_1(M)} \right) : 
\left( {M_3(M)\over \alpha_3(M)}- {M_1(M)\over \alpha_1(M)} \right) =
2 :3.
\label{ratiochi2}
\end{equation}
At the $M$-scale, the gauge couplings are
$\alpha^{-1}_{1,2,3}\approx 30.7, 25.6, 22.9$. 
Equations~(\ref{ratiochi}) and (\ref{ratiochi2}) are 
one-loop RGE invariant so they approximately 
hold at the electroweak scale as well.

From the boundary condition
Eq.~(\ref{gm0}) and the RGE evolution, we obtain a gaugino 
mass relation at the $M_Z$-scale, compared with
those in the mGMSB or mSUGRA models,
\[ m_{\tilde g} : m_{\tilde\chi^0_2,\tilde\chi^\pm_1} :
m_{\tilde\chi^0_1} \approx 
\left\{ \begin{array}{rl}
         22 : 6 : 1 & \mbox{for $\nphi=\nsig=1$},\\ 
          6 : 2 : 1 &  \mbox{for mGMSB or mSUGRA}.
        \end{array}
\right. \]
In our scenario, $\tilde\chi^0_1$ is basically $\tilde B$, 
and $\tilde\chi^\pm_1,\tilde\chi^0_2$ 
are $\widetilde W^\pm,\widetilde Z^0$. 

The additional contribution from the adjoint messengers to
scalar masses generally yields heavier scalars in this
model. However, the masses of the right-handed 
sleptons do not receive any correction from them. 
Equation~(\ref{gm1}) gives a mass relation for the
sfermion soft masses at the $M$-scale
\begin{eqnarray}
&&\qquad\qquad m^2_{\tilde Q} : m^2_{\tilde U} : m^2_{\tilde D} : 
m^2_{\tilde L, H_u, H_d} : m^2_{\tilde E}\ \approx\  \nonumber\\
&&(15.6 \nsig + 5.8 \nphi) : (12 \nsig + 4.5 \nphi)
: (12 \nsig + 4 \nphi) : (3.6 \nsig + 2 \nphi) : \nphi\ .
\label{ratiosq}
\end{eqnarray}
Squark masses receive a large contribution from the 
gluino soft mass $M_3$ via the RGE evolution. At the $M_Z$-scale,
we obtain a very simple relation among the masses
for the first two generation squarks and sleptons, 
also compared with that in the mGMSB,
\[ m^{}_{{\tilde u}_{L,R}, {\tilde d}_{L,R}} : 
   m^{}_{{\tilde\nu}, {\tilde e}_L} : 
   m^{}_{{\tilde e}_R} \approx 
\left\{ \begin{array}{rl}
        9 : 3 : 1 & \mbox{for $\nphi=\nsig=1$},\\
        6 : 2 : 1 & \mbox{for mGMSB with $M\approx 100$ TeV}.
        \end{array}
\right. \]

In the case $\nphi=\nsig=1$, the ratios of the 
$M_3, M_2, M_1$ to $m_{\tilde E}$ turn out to
be $4.7:3.2:0.9:1$ at the $M$-scale. 
Evolving to the $M_Z$-scale,
we find the mass ratio 
\[ m_{{\tilde e}_R} : m_{\tilde\chi_1^0}\approx
\left\{ \begin{array}{ll}
        2.4  & \mbox{for $\nphi=\nsig=1$},\\
        1.4 & \mbox{for mGMSB with $M\approx 100$ TeV}.
        \end{array}
\right. \]
Since the adjoint messengers carry no $U_Y(1)$ charge, 
they do not contribute to $m^{}_{\tilde\chi^0_1}$ 
and $m_{{\tilde \ell}_R}$. The above mass difference comes 
entirely from the RGE evolution
at the two different messenger scales, namely $10^5$ GeV 
for mGMSB and $10^{14}$ GeV for our model. 
This mass ratio thus provides a direct measure
in extracting the underlying messenger mass scale.

\begin{figure}[tb]
\epsfysize=2.5in
\epsffile[0 270 380 515]{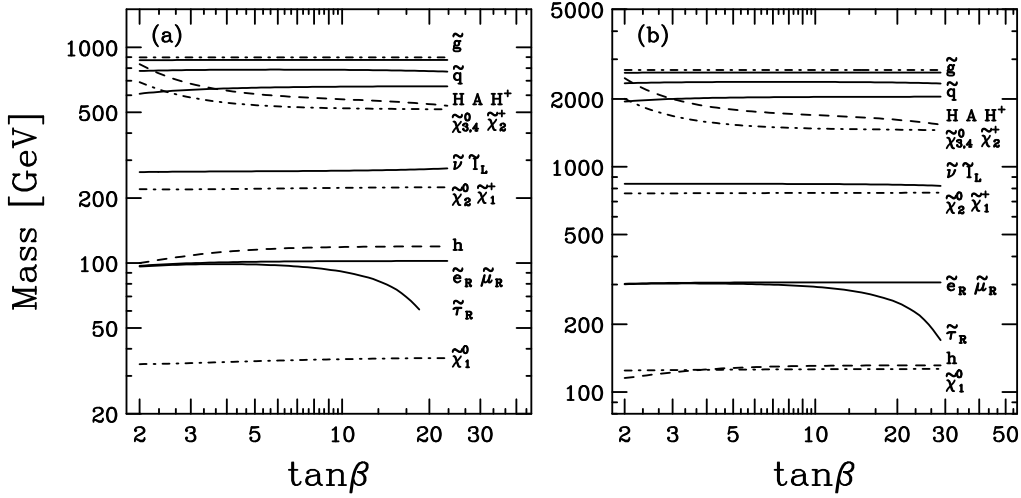}
\begin{center}
\parbox{5.5in}{
\caption[]{Predicted sparticle mass spectra at the 
electroweak scale versus $\tan\beta$ for (a) 
$\La=30$ TeV and (b) $\La=100$ TeV, where $\mu<0$, 
$M=10^{14}$ GeV and $\nphi=\nsig=1$ are assumed.
\label{masses}}}
\end{center}
\end{figure}

If $\La \sim {\cal O}$($30-100$ TeV),
the sparticles can have a desirable mass 
spectrum of ${\cal O}$(100 GeV). The 
predicted sparticle mass spectra at the $M_Z$-scale 
are presented in Fig.~\ref{masses}, with (a) for
$\La=30$ TeV and (b) for $\La=100$ TeV, where we have
implemented the two-loop evolution for the RGEs,
and have properly taken the radiative electroweak 
symmetry breaking into account.
The mass spectrum exhibits apparent hierarchy 
relations as noted in the previous discussions.
The exceptions happen for the third generation 
squarks and sleptons, where the Yukawa contributions
in the RGE evolution are also important. In particular,
the lightest tau-slepton $\tilde\tau_R^{}$
 could be significantly 
lighter than other scalar particles, 
as also known from the mGMSB models.
The gluino happens to be the heaviest, and next come
the squarks. There are three curves for the squarks right
below the gluino mass. The lower two are for the lighter 
top-squarks $\tilde t_{1,2}$, 
and the upper one is for the other nearly 
degenerate squarks.

From Eq.~(\ref{mscale}),
we estimate that $F_S\approx \La M \approx 10^{18}-10^{19}$ (GeV)$^2$.
This implies a relatively heavy gravitino
\begin{equation}
m_{3/2} =  {F_S\over \sqrt 3\ M^*_{\rm Pl}} \approx 1-10\ {\rm GeV},
\label{gold}
\end{equation}
where $M^*_{\rm Pl}=2.4\times 10^{18}$ GeV is the reduced Planck scale.
This rather heavy stable gravitino may
form significant amount of dark matter. As long as
the reheating temperature after the inflation does not exceed about
$T_{\rm RH}\approx 10^8$ GeV, their relic density would 
not be too high to overclose the Universe \cite{gmm}. 
It has been argued that in realistic string theories, 
the light moduli remnants 
(like the gravitino here) may distort the observed
X-ray spectrum by radiative decay through gravitational
effects \cite{modulidec}. It turns out that it would not 
destroy the observed spectrum if 
$m_{3/2}\ge {\cal O}(1$ GeV). On the other hand,
if $m_{3/2}\ge 10$ GeV, the scalar mass universality might be 
violated by the large gravitational contribution and 
the unwanted flavor-changing neutral currents (FCNC) 
may be reintroduced \cite{DN}. Although our estimate
on $m_{3/2}$ is essentially on the safe side for the FCNC problem, 
this consideration may serve as a criterion for favoring
$\nsig =1$, as noted earlier. 

Typically, the next-to-lightest SUSY particle (NLSP) in GMSB models
is either $\tilde \chi^0_1$ or $\tilde \tau^{}_R$ 
(for large $\tan\beta$ and higher $\nphi$). 
With such a heavy gravitino, the NLSP would be very
long-lived, with a decay length
much larger than the size of the detectors.
%
%
The NLSP would appear to be stable in the collider 
environment. This would imply that
the standard missing-energy technique should be applicable
for the searches if $\tilde \chi^0_1$ is the NLSP, while 
a heavy charged track in the detector would be the 
signal for $\tilde \tau_R$ as the NLSP.

The mass spectrum scales linearly with the parameter $\La$.
For $\La=30$ TeV, we have a relatively light mass spectrum,
which can be accessed by next generation collider experiments,
while for $\La=100$ TeV, most sparticles are probably not
easy to be produced except for $h$, $\tilde \chi^0_1$ and 
$\tilde \ell_R^{}$. One can also interpolate
the mass spectrum for the $\La$-parameter in between.

\vskip 0.1in
{\noindent \bf{Further Remarks:}}~
Concerning the schemes in preserving the gauge coupling
unification beyond the MSSM particle contents
at a high energy scale, we would like to re-emphasize 
that adding in more states in complete representations
of the GUT group would automatically keep the unification
without changing the GUT scale; while introducing the
(matching pair) adjoint representations of the SM gauge 
group would also keep the unification, but generally
shift the GUT scale, depending on the mass
threshold of the adjoint states. This is applicable
even beyond the specific model under discussion,
namely, aiming only at $M_\str$. In principle, the
gauge coupling unification could occur at a different
scale, regardless of the heterotic string prediction 
$M_\str$. However, our idea with the adjoint states
beyond the MSSM as messengers at a high $M$-scale, 
which help preserve the unification, 
can be tested against the distinctive
sparticle spectrum prediction by future collider
experiments.

Regarding the origin of the adjoint messengers
$\sige$ and $\sigt$, we noted that they may be identified
as the remnants resulting from certain realistic
string models as continuous moduli \cite{bfy}. 
In fact, although highly model-dependent, there are
often other vector-like representations which could
provide the $\Phi_5+{\overline \Phi}_5$ states
in $SU(5)$ as well \cite{strgut}. Along the similar line,
an attempt \cite{alon} has been made in which the messenger
sector consists solely of color triplets, arising
from the Wilson-line breaking of unifying non-Abelian
gauge symmetries in string models. However, 
this model predicted a very light sparticle
spectrum that has been excluded by the LEP-II
experiments.

Although typical GMSB models are generally
lack of satisfactory
cold dark matter candidates \cite{colddm}, a stable
heavy particle associated with our messenger sector
may provide a superheavy dark matter candidate
with $M={\cal O}(10^{14}$ GeV) \cite{rockyetal}.
More investigation in this regard is needed 
before drawing a conclusion. 

\vskip 0.1in
{\noindent \bf {Conclusion:}}~
We have introduced the adjoint messengers $\sige$ and $\sigt$ 
for gauge-mediation of SUSY breaking. 
These new messengers lift the
gauge coupling unification to the string scale at 
${\cal O}(5\times 10^{17}$ GeV) if their masses are at 
${\cal O}(10^{14}$ GeV). This proposed ``marriage'' may
have profound implication:
some remnant states in certain realistic string models
may serve as the messengers for gauge-mediation of SUSY
breaking. The model is highly predictive and restrictive.
The predicted SUSY mass spectrum 
at the electroweak scale is significantly different 
from those in other GMSB and mSUGRA models, resulting 
in experimentally testable robust mass relations. 
The gravitino mass is predicted to
be approximately $1-10$ GeV. Consequently,
the NLSP appears to be very long-lived and would only
decay outside the detector in the collider environment.
The very heavy stable particle associated with the
messenger sector may also provide a 
superheavy dark matter candidate.

\vskip 0.1in
{\noindent \bf Acknowledgment}

T.Y. would like to thank the Phenomenology Institute
of the UW-Madison for hospitality when this work was
initiated. This work was supported in part by a DOE 
grant No. DE-FG02-95ER40896, and in part by the 
Wisconsin Alumni Research Foundation.

\end{document}